\documentclass[epj]{svjour}
\usepackage{graphics}

\usepackage{graphicx}
\usepackage[dvips]{epsfig}
\usepackage{subfigure}

\newcommand{\dd}{\mbox{\rm d}}
\newcommand{\dpce}{\mbox{$dp\to \{pp\}n$}}

\newcommand{\dpcepol}{$d\hspace{-3.2mm}\stackrel{\to}{\phantom{p}}\hspace{-1.4mm}p\to\{pp\}n$}

\newcommand{\half}{\mbox{${\textstyle \frac{1}{2}}$}}

\newcommand{\Szero}{\mbox{$^{1\!}S_0$}}

\begin{document}
\title{Neutron--proton charge--exchange amplitudes at 585~MeV}
\author{D.~Chiladze\inst{1,2,3}\and
 J.~Carbonell\inst{4}\and
 A.~Dzyuba\inst{5} \and
 S.~Dymov\inst{6,7}\and
 V.~Glagolev\inst{8}\and
 M.~Hartmann\inst{2,3}\and
 A.~Kacharava\inst{2,3}\and
 I.~Keshelashvili\inst{1,9}\and
 A.~Khoukaz\inst{10}\and
 V.~Komarov\inst{6}\and
 P.~Kulessa\inst{11}\and
 A.~Kulikov\inst{6}\and
 N.~Lomidze\inst{1}\and
 G.~Macharashvili\inst{1,6}\and
 Y.~Maeda\inst{12}\and
 D.~Mchedlishvili\inst{1}\and
 T.~Mersmann\inst{10}\and
 S.~Merzliakov\inst{2,3,6}\and
 M.~Mielke\inst{10}\and
 S.~Mikirtychyants\inst{2,3,5}\and
 M.~Nekipelov\inst{2,3}\and
 M.~Nioradze\inst{1}\and
 H.~Ohm\inst{2,3}\and
 F.~Rathmann\inst{2,3}\and
 H.~Str\"oher\inst{2,3}\and
 M.~Tabidze\inst{1}\and
 S.~Trusov\inst{13}\and
 Yu.~Uzikov\inst{6}\and
 Yu.~Valdau\inst{5}\and
 C.~Wilkin\inst{14}\thanks{e-mail: cw@hep.ucl.ac.uk} }
\authorrunning{D.~Chiladze \emph{et.al.}}
%
\institute{High Energy Physics Institute, Tbilisi State University, 0186
Tbilisi, Georgia
\and  Institut f\"ur Kernphysik, Forschungszentrum J\"ulich GmbH, 52425
J\"ulich, Germany
\and J\"ulich Centre for Hadron Physics, 52425 J\"ulich, Germany
\and  Laboratoire de Physique Subatomique et de Cosmologie, 38026
Grenoble, France
\and High Energy Physics Department, Petersburg Nuclear Physics
Institute, 188350 Gatchina, Russia
\and Laboratory of Nuclear Problems, JINR, 141980 Dubna, Russia
\and  Physikalisches Institut II, Universit\"at Erlangen--N\"urnberg,
91058 Erlangen, Germany
\and Laboratory of High Energies, JINR, 141980 Dubna, Russia
\and Department of Physics, University of Basel, Klingelbergstrasse
82, 4056 Basel, Switzerland
\and Institut f\"ur Kernphysik, Universit\"at M\"unster, 48149
M\"unster, Germany
\and H.~Niewodnicza\'{n}ski Institute of Nuclear Physics PAN, 31342
Krak\'{o}w, Poland
\and Research Center for Nuclear Physics, Osaka University, Ibaraki,
Osaka 567-0047, Japan
\and Institut f\"ur Kern- und Hadronenphysik,
Forschungszentrum Rossendorf, 01314 Dresden, Germany
\and Physics and Astronomy Department, UCL, Gower Street, London,
WC1E 6BT, UK}

\date{Received: \today / Revised version:}
\abstract{The differential cross section and deuteron analysing
powers of the \dpcepol\ charge--exchange reaction have been measured
with the ANKE spectrometer at the COSY storage ring. Using a deuteron
beam of energy 1170~MeV, data were obtained for small momentum
transfers to a $\{pp\}$ system with low excitation energy. A good
quantitative understanding of all the measured observables is
provided by the impulse approximation using known neutron--proton
amplitudes. The proof of principle achieved here for the method
suggests that measurements at higher energies will provide useful
information in regions where the existing $np$ database is far less
reliable.
} 
\maketitle

%
%

\section{Introduction}
\label{intro}

An understanding of the $NN$ interaction is fundamental to the whole
of nuclear and hadronic physics. The database on proton--proton
elastic scattering is enormous and the wealth of spin--dependent
quantities measured has allowed the extraction of $NN$ phase shifts
in the isospin $I=1$ channel up to a beam energy of at least
2~GeV~\cite{SAIDNN}. The situation is far less advanced for the
isoscalar channel where the much poorer neutron--proton data only
permit the $I=0$ phase shifts to be evaluated up to at most 1.3~GeV
but with significant ambiguity above about 800~MeV~\cite{SAIDNN}. The
data on which such an analysis is based come from many facilities and
it is incumbent on a laboratory that can make a significant
contribution to the communal effort to do so.

It has recently been argued that, even without measuring triple--spin
observables, a direct amplitude reconstruction of the neutron--proton
backward scattering amplitudes might be possible with few ambiguities
provided that experiments on the deuteron are included~\cite{Lehar}.
This work studied in detail the ratio of the forward charge--exchange
cross section of a neutron on a deuterium target to that on a
hydrogen target,
\begin{equation}
R_{np}(0)=\frac{\dd\sigma(nd\to pnn)/\dd{t}} {\dd\sigma(np\to
pn)/\dd{t}}, \label{Rnp}
\end{equation}
where $t$ is the four--momentum transfer between the initial neutron
and final proton.

Due to the Pauli principle, when the two final neutrons are in a
relative $S$--wave their spins must be antiparallel and the system is
in the \Szero\ state. Under such circumstances the $nd\to p\{nn\}$
reaction involves a spin flip from the $S=1$ of the deuteron to the
$S=0$ of the dineutron and hence is dependent on the $np$
spin-isospin-flip amplitudes. If the data are summed over all
excitation energies of the $nn$ system, then the Dean closure sum
rule allows one to deduce from $R_{np}$ the fraction of
spin--dependence in the $pn$ charge--exchange amplitudes~\cite{Dean}.
Such measurements have now been carried out up to
2~GeV~\cite{Sharov}.

However, Bugg and Wilkin~\cite{BW} have shown that much more
information on the $np$ charge--exchange amplitudes can be extracted
by using a polarised deuteron beam or target and studying the
charge--symmetric \dpcepol\ reaction. To achieve the full benefit,
the excitation energy $E_{pp}$ in the final $pp$ system must be kept
low. Experiments from a few hundred MeV up to
2~GeV~\cite{Ellegaard,Kox} have generally borne out the theoretical
predictions and have therefore given hope that such experiments might
provide valuable data on the amplitudes in the small momentum
transfer region.

The ANKE collaboration is embarking on a systematic programme to
measure the differential cross section and analysing powers of the
\dpcepol\ reaction up to the maximum energy of the COSY accelerator
of 1.15~GeV per nucleon, with the aim of deducing information on the
$np$ amplitudes~\cite{SPIN}. Higher energies per nucleon will be
achieved through the use of a deuterium target. Spin correlations
will also be studied with a polarised beam and target~\cite{PIT}. However, for
these to be valid objectives, the methodology has to be checked in a
region where the neutron--proton amplitudes are reasonably well
known.

The first evaluation of the analysing powers of the \dpcepol\
reaction at $T_d=1170$~MeV reported in Ref.~\cite{Chiladze0} largely
agrees with theoretical expectations. It is the purpose of the
present work to refine this analysis and to establish also the cross
section normalisation. In this way the magnitudes of individual
charge--exchange amplitudes could be tested and not merely their
ratios.

Although the impulse approximation description of the \dpce\ reaction
is to be found in several papers~\cite{BW,Carbonell}, a brief
resum\'{e} is presented in section~\ref{Impulse} for the ideal case
where the final $pp$ system is in a pure \Szero\ state. The general
layout and capabilities of the ANKE facility are described in
section~\ref{expset}. The normalisation of the charge--exchange cross
sections is achieved relative to the quasi--free $dp\to p_{\,{\rm
sp}} d\pi^0$ reaction; the detection of a spectator proton in
coincidence with the deuteron, produced \emph{via} $np\to d\pi^0$,
closely matches the acceptance for the two charge-exchange protons.
These measurements are described in section~\ref{Illuminate}. The
luminosity obtained by comparing the results with those in the
literature allows the unpolarised \dpce\ differential cross section
to be determined, as shown in section~\ref{dce}. Some check on the
luminosity could be provided through the study of elastic
deuteron--proton scattering, though there are larger uncertainties in
the relevant World database.

In order that the deuteron analysing powers can be measured, the
value of the polarisation has to be established for each of the
modes of the ion source used in the preparation of the beam. By using the charge--exchange
data themselves, it has proved possible to reduce the statistical error
bars residing in the earlier analysis~\cite{Chiladze0,Chiladze1} and
this is explained in section~\ref{polarisation}. The evaluation of
the \dpcepol\ analysing powers as a function of the momentum transfer
in two bins of $E_{pp}$ is the subject treated in section~\ref{Ap}.

Since both the cross section and two tensor analysing powers at
585~MeV per nucleon agree with theoretical predictions based upon
reliable neutron--proton phase--shift analysis, this gives us
confidence that the methods used here can be extended to higher
energies where much less is known about the $np$ elastic amplitudes.
The possibilities of such work are discussed in
section~\ref{Conclusions}.
%
%
\section{Impulse approximation dynamics}
\label{Impulse}\setcounter{equation}{0}

Bugg and Wilkin studied the cross section and deuteron analysing
powers of the \dpcepol\ reaction within the impulse
approximation~\cite{BW} and their results were refined through the
use of more realistic low energy nucleon--nucleon interactions in
Ref.~\cite{Carbonell}. In this approach it is assumed that the
driving mechanism is a quasi--free $(p,n)$ charge exchange on the
neutron in the deuteron. The resulting matrix element is then
proportional to that for $pn\to np$ times a form factor that depends
upon the deuteron and $pp$ wave functions and the momentum transfer
$\vec{q}$. There is a strong interplay between the spin dependence of
the $np$ amplitudes and the polarisation of the initial deuteron and
this leads to very significant deuteron tensor analysing powers.
However, it is crucial to note that these analysing powers tend to
have opposite signs for spin--singlet and spin--triplet $pp$ final
states~\cite{BW}. As a consequence, the sizes of the resulting
effects will depend strongly on the limits placed upon the $pp$
excitation energy in order to isolate the \Szero\ state. We here
merely present explicitly the formalism for a pure $S$-wave state
while recognising that the detailed comparison of data with theory
requires a much more thorough numerical evaluation of the full
model~\cite{Carbonell}.

The charge--exchange amplitude of the elementary $np\to pn$
scattering may be written in terms of five scalar amplitudes in the
cm system as
\begin{eqnarray}
\nonumber f_{np}&=&\alpha(q) +i\gamma(q)
(\vec{\sigma}_{1}+\vec{\sigma}_{2})\cdot\vec{n} +\beta(q)
(\vec{\sigma}_{1} \cdot {\bf n})(\vec{\sigma}_{2}\cdot\vec{n})\\
&&+\delta(q)(\vec{\sigma}_{1}\cdot\vec{m})(\vec{\sigma}_{2}\cdot\vec{m})
+\varepsilon(q)(\vec{\sigma}_{1}\cdot\vec{l})(\vec{\sigma}_{2}\cdot\vec{l}),
 \label{fpn}
\end{eqnarray}
where $\vec{\sigma}_{1} $ is the Pauli matrix between the initial
neutron and final proton, and the reverse for $\vec{\sigma}_{2}$. As
stressed in Ref.~\cite{Lehar}, $\alpha$ is the spin--independent
amplitude between the initial neutron and final proton, $\gamma$ is a
spin--orbit contribution, and $\beta$, $\delta$, and $\varepsilon$
are the spin--spin terms. The one--pion--exchange pole is contained
purely in the $\delta$ amplitude and this gives rise to its very
rapid variation with momentum transfer, which influences very
strongly the deuteron charge--exchange observables.

The orthogonal unit vectors are defined in terms of the initial
neutron ($\vec{K}$) and final proton ($\vec{K'}$) momenta;
\begin{equation}
\vec{n}=\frac{\vec{K}\times\vec{K'}}{|\vec{K}\times\vec{K'}|},~~
\vec{m}=\frac{\vec{K'}-\vec{K}}{|\vec{K'}-\vec{K}|},~~
\vec{l}=\frac{\vec{K'}+\vec{K}}{|\vec{K'}+\vec{K}|}\,\cdot
\end{equation}
The amplitudes are normalised such that the elementary $np\to pn$
differential cross section has the form
\begin{equation}
\label{secpn} \left(\frac{\dd\sigma}{\dd t}\right)_{\!\!np\to pn}=
|\alpha(q)|^{2}+|\beta(q)|^{2}+2|\gamma(q)|^{2}
+|\delta(q)|^{2}+|\varepsilon(q)|^{2}\,.
\end{equation}

In impulse approximation the deuteron charge--exchange amplitude to
the \Szero\ state depends only upon the spin--dependent parts of
$f_{np}$~\cite{BW}. The form factor describing the transition from a
deuteron to a \Szero --state of the final $pp$ pair contains two
terms
\begin{eqnarray}
\nonumber
S^{+}(k, \half q)&=&F_{S}(k, \half q)+\sqrt{2}F_{D}(k, \half q)\,,\\
\label{FormF}
S^{-}(k, \half q)&=&F_{S}(k, \half q)-F_{D}(k, \half q)/\sqrt{2}\,.
\end{eqnarray}
Here $\vec{q}$ is the three--momentum transfer between the proton and
neutron which, for small $E_{pp}$, is related to the four--momentum
transfer by $t=-\vec{q}^2$.

The $S^{+}$ and $S^{-}$ denote the longitudinal ($\lambda=0$) and
transverse ($\lambda=\pm1$) form factors, where $\lambda$ is the
spin--projection of the deuteron. The matrix elements $F_S$ and $F_D$
can be expressed in terms of the $S$-- and $D$--state components of
the deuteron wave function $u(r)$ and $w(r)$ and the
$pp~(\Szero)$--scattering wave function $\psi^{(-)}_k(r)$ as
\begin{eqnarray}
F_{S}(k, \half q)&=&\langle\psi _{k}^{(-)}|j_{0}(\half qr)|u\rangle\,,\\
F_{D}(k, \half q)&=&\langle\psi _{k}^{(-)}|j_{2}(\half qr)|w\rangle.
\end{eqnarray}
Here $\vec{k}$ is the $pp$ relative momentum, corresponding to an
excitation energy $E_{pp}=\vec{k}^2/m$, where $m$ is the proton mass.
We denote the ratio of the transition form factors by
\begin{equation}
\label{ratio} \mathcal{R}=\left.{S^{+}(k,\half
q)}\right/{S^{-}(k,\half q)}
\end{equation}
and the combination of modulus--squares of amplitudes by
\begin{equation}
\label{idce} I=|\beta(q)|^{2}+|\gamma(q)|^{2}+|\varepsilon(q)
|^{2}+|\delta(q)|^{2}|\mathcal{R}|^{2}.
\end{equation}
Impulse approximation applied to $dp\to \{pp\}_{^1S_0}n$ then leads
to the following predictions for the differential cross section and
deuteron spherical analysing powers~\cite{BW,Carbonell}:
\begin{eqnarray}
\nonumber \frac{\dd^{4}\sigma}{\dd{t}\,\dd^{3}k}&=&
\left.I\left[S^{-}(k,\half q)\right]^2\!\!\right/\!3,\\
\nonumber
I\,it_{11}&=&0\:,\\
\nonumber
I\,t_{20}&=&\left[|\beta(q)|^2+|\delta(q)|^2|\mathcal{R}|^2 -2|\varepsilon(q)|^2+|\gamma(q)|^2\right]/\sqrt{2}\:,\\
I\,t_{22}&=&\sqrt{3}\left[|\beta(q)|^2-|\delta(q)|^2|\mathcal{R}|^2+|\gamma(q)|^2\right]/2\:.
\label{impulse}
\end{eqnarray}

In this \Szero\ limit, a measurement of the differential cross
section, $t_{20}$, and $t_{22}$ would allow one to extract values of
$|\beta(q)|^2+|\gamma(q)|^2$, $|\delta(q)|^2$, and
$|\varepsilon(q)|^2$ for small values of the momentum transfer
$\vec{q}$ between the initial proton and final neutron. However, even
if a sharp cut of 1~MeV is placed upon $E_{pp}$, there still remain
small contributions from proton--proton $P$--waves that dilute the
analysing power signal. Such effects must be included in any analysis
aimed at providing quantitative information on the neutron--proton
amplitudes~\cite{Carbonell}.

One way of reducing the dilution of the tensor analysing powers by
the $P$--waves is by imposing a cut on the angle $\theta_{qk}$
between the momentum transfer $\vec{q}$ and the relative momentum
$\vec{k}$ between the two protons. When these two vectors are
perpendicular, impulse approximation does not allow the excitation of
odd partial waves in the $pp$ system~\cite{BW} and this is in
agreement with available experimental data~\cite{Kox}.

In terms of the charge--exchange amplitudes, the Dean sum
rule~\cite{Dean} for the ratio of the forward $nd\to pnn$ to $np\to
pn$ cross sections of Eq.~(\ref{Rnp}) becomes
\begin{equation}
\label{Dean} R_{np}(0) =
\frac{2}{3}\left[\frac{2|\beta(0)|^2+|\varepsilon(0)|^2}
{|\alpha(0)|^2+2|\beta(0)|^2+|\varepsilon(0)|^2}\right]\cdot
\end{equation}
The same result should, of course, hold for $dp\to ppn$, which is the reaction
studied at ANKE.

%
%
\section{The experimental facility}
\label{expset}\setcounter{equation}{0}

\begin{figure}[htb]
\begin{center}
\resizebox{0.46\textwidth}{!}{%
\includegraphics{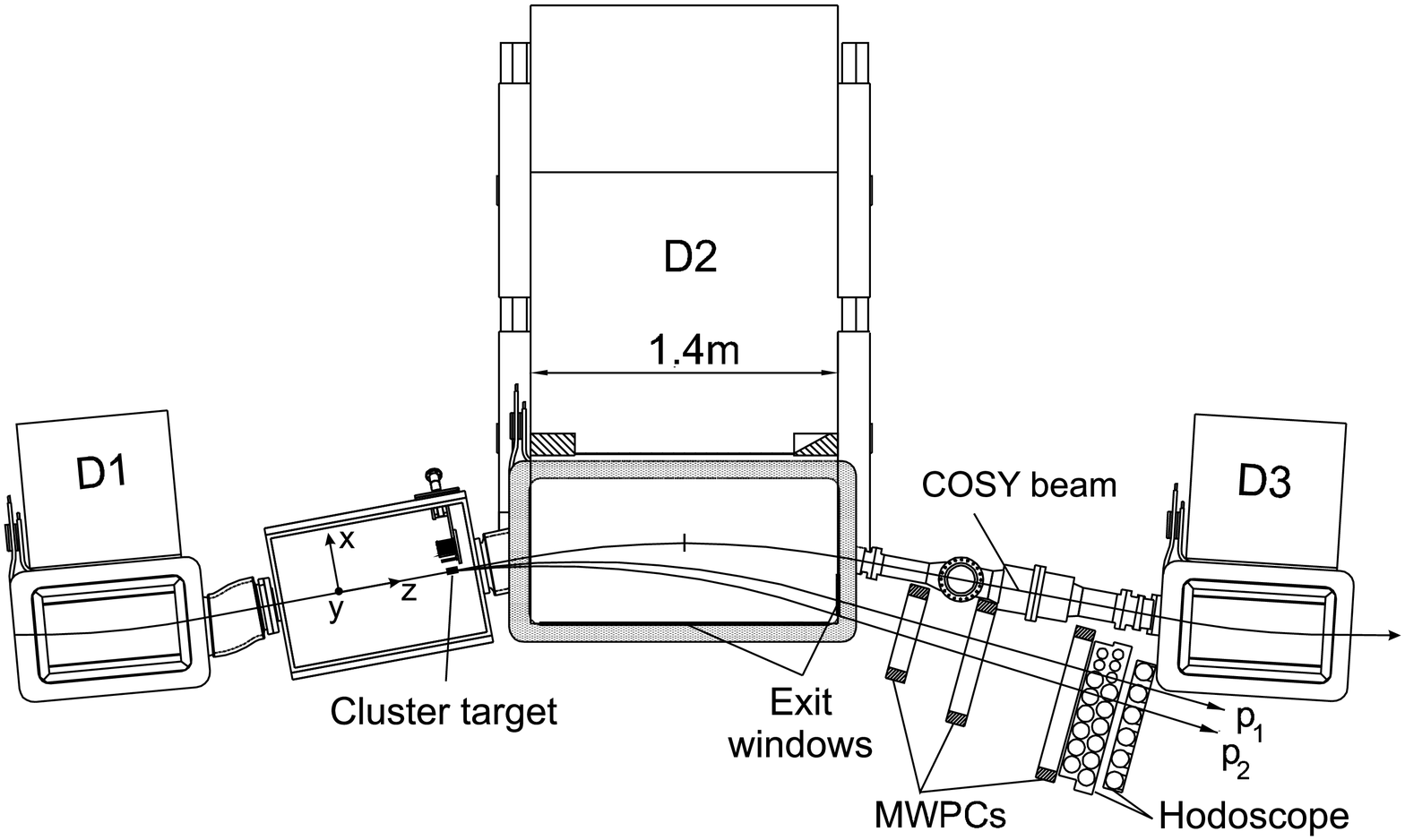}}
\caption{Top view of the ANKE experimental set--up, showing the
positions of the three dipole magnets D1, D2, and D3. The hydrogen
cluster-jet injects target material vertically downwards. The Forward
Detector (FD) consists of three MWPCs and a hodoscope composed of
three layers of scintillation counters.} \label{Wig01}
\end{center}
\end{figure}

The COSY COoler SYnchrotron of the Forschungszentrum J\"ulich is
capable of accelerating and storing protons and deuterons with
momenta up to 3.7~GeV/c~\cite{Mai97}. The ANKE magnetic spectrometer
of Fig.~\ref{Wig01} used for the deuteron charge-exchange study is
located at an internal target position that forms a chicane in the
storage ring. Although ANKE contains several detection
possibilities~\cite{ANKE}, only those of the Forward Detector (FD)
system were used to measure the two fast protons from the \dpce\
charge exchange~\cite{Chiladze0}, as well as the products associated
with the calibration reactions. The FD consists of multiwire chambers
for track reconstruction and three layers of a scintillation
hodoscope that permit time--of--flight and energy--loss
determinations~\cite{Dymov}. The measurements were carried out using
a polarised deuteron beam and a hydrogen cluster--jet
target~\cite{Khoukaz}. The main trigger used in the experiment consisted of a
coincidence of different layers in the hodoscope of the FD.

\begin{figure}[htb]
\begin{center}
\resizebox{0.44\textwidth}{!}{%
\includegraphics{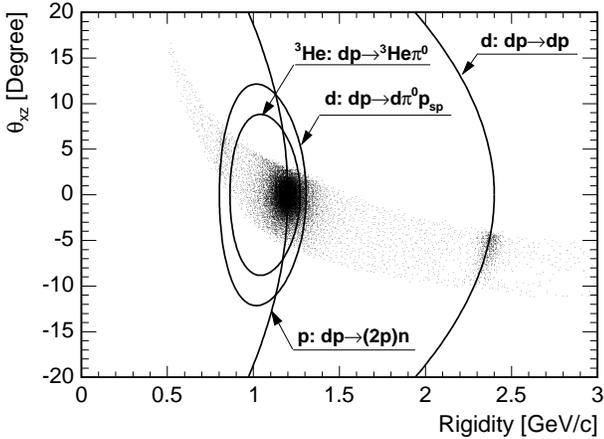}}
\caption{ANKE experimental acceptance for four nuclear reactions of
interest at a deuteron momentum of $p_d=2400$~MeV/c.} \label{Wig02}
\end{center}
\end{figure}

Figure~\ref{Wig02} shows the experimental acceptance of ANKE for
single particles at $T_d=1170$~MeV in terms of the laboratory
production angle in the horizontal plane and the magnetic rigidity.
The kinematical loci for various nuclear reactions are also
illustrated. In addition to the protons from the deuteron charge
exchange \dpce, of particular interest are the deuterons produced in
the quasi-free $dp\to p_{\rm sp}d\pi^0$ reaction with a fast
spectator proton, $p_{\rm sp}$. It is important to note that these
spectators, as well as those from the deuteron breakup, $dp \to
p_{\rm sp}pn$, have essentially identical kinematics to those of the
charge--exchange protons. As a consequence, the $(d,2p)$ reaction can
only be distinguished from other processes yielding a proton
spectator by carrying out coincidence measurements. Deuterons
elastically scattered at small angles are well separated from the
other particles in Fig.~\ref{Wig02}.

%
%
\section{Luminosity measurements}
\label{Illuminate}

%
%
\subsection{Quasi--free pion production}
\label{pizero}\setcounter{equation}{0}

Both the fast deuteron and the spectator proton, $p_{\rm sp}$, from
the $dp\to p_{\rm sp}d\pi^0$ reaction have momenta that are very
similar to those of the two protons in the \dpce\ reaction. Any error
in the estimation of the two-particle acceptance will therefore tend
to cancel between the two reactions. Interpreting the data in terms
of quasi--free pion production, $np\to d\pi^0$, the counting rates
for the $dp\to p_{\rm sp}d\pi^0$ reaction will allow a useful
evaluation of the luminosity to be made.

\begin{figure}[hbt]
\begin{center}
\resizebox{0.44\textwidth}{!}{\includegraphics{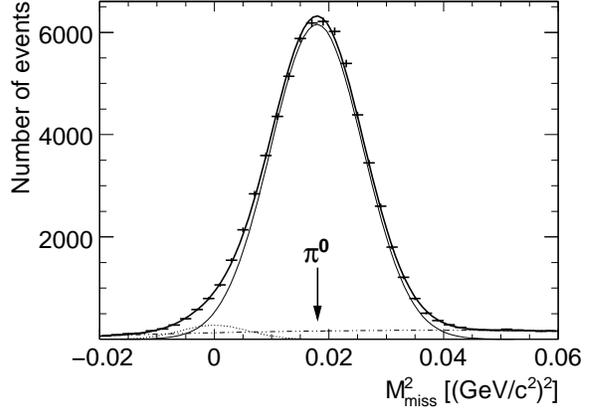}}
\caption{Missing--mass--squared distribution for the $dp\to p_{\rm
sp}d X$ reaction for the fast deuteron branch of the kinematics. A
fit to the data with a Gaussian and a constant background is
indicated. Some of the event excess for $M_{\rm miss}^2\approx 0$
might be associated with the quasi--free $np\to d\gamma$ reaction.}
\label{Wig03}
\end{center}
\end{figure}

The first step in extracting quasi--free $dp\to p_{\rm sp}d\pi^0$
candidates from the data is to choose two--track events using the
MWPC information. The momentum vectors were determined from the
magnetic field map of the spectrometer, assuming a point--like source
placed in the centre of the beam--target interaction region. The
potential $dp\to p_{\rm sp}d X$ events can be clearly identified by
studying the correlation of the measured time difference between the
two hits on the hodoscope with that calculated on the basis of the
distances from the target and the two momenta~\cite{Chiladze1}. In
order to ensure that the kinematics are similar to the two protons
from the charge exchange at low $E_{pp}$, a cut is made on the
difference between the momenta of the assumed proton and deuteron of
$\Delta p < 175$~MeV/c. An analogous cut was placed upon the
simulation of the acceptance.

The $dp\to p_{\rm sp}d\pi^0$ identification is completed by studying
the missing mass of the reaction with respect to the final $dp$ pair,
as shown in Fig.~\ref{Wig03}. The $\Delta p$ cut means that only
events corresponding to the forward deuteron branch are presented
here. As is seen from Fig.~\ref{Wig02}, these ones have similar
acceptance to that of the \dpce\ reaction. The data show a very
prominent $\pi^0$ peak though there is evidence for background on the
low $M_{\rm miss}^2$ side, some of which might be arise from the
quasi--free $np \to d\gamma$ reaction.

\begin{figure}[hbt]
\begin{center}
\resizebox{0.44\textwidth}{!}{\includegraphics{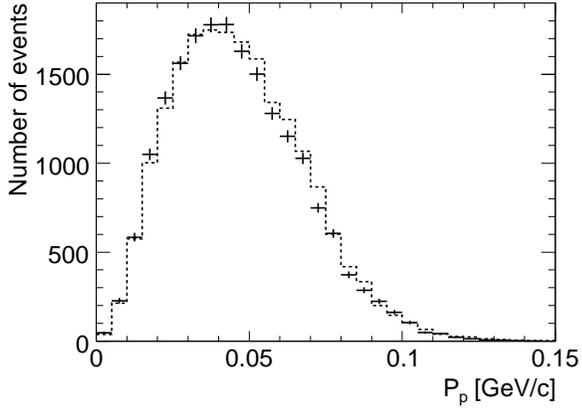}}
\caption{Momentum distribution of the fast proton from the $dp\to
p_{\rm sp} d\pi^0$ reaction transformed into the rest frame of the
initial deuteron (histogram). The simulation (crosses) uses the Fermi
momentum distribution obtained from the Paris deuteron wave
function~\cite{Paris}. Only data with $p_{\rm sp} < 60$~MeV/c were
used in the evaluation of the luminosity.} \label{Wig04}
\end{center}
\end{figure}

\begin{figure}[htb]
\begin{center}
\resizebox{0.44\textwidth}{!}{\includegraphics{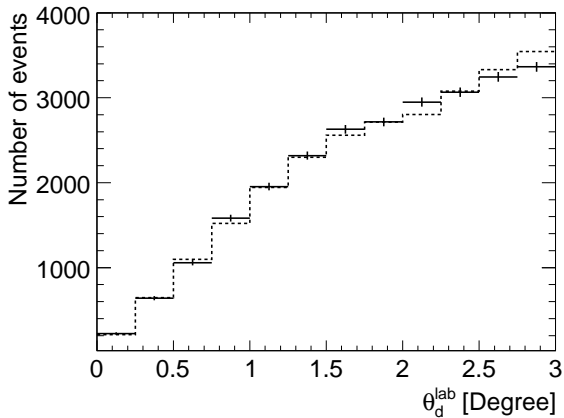}}
\caption{Corrected numbers of counts of quasi--free $np\to d\pi^0$
events in $0.25^{\circ}$ bins (crosses). The histogram represents the
prediction of the $pp\to d\pi^+$ differential cross section taken
from the SAID program~\cite{SAIDpid}. After taking an isospin factor
of two into account, scaling the simulation to agree with the
experimental points allows the luminosity to be evaluated.
\label{Wig05}}
\end{center}
\end{figure}

To confirm the spectator hypothesis, a Monte Carlo simulation has
been performed within PLUTO~\cite{PLUTO} using the Fermi momentum
distribution evaluated from the Paris deuteron wave
function~\cite{Paris}. As is clear from Fig.~\ref{Wig04}, the data
are completely consistent with quasi-free production on the neutron
leading to a spectator proton. However, in order to reduce further
possible contributions from multiple scattering and other mechanisms,
only events with $p_{\rm sp} < 60$~MeV/c were retained for the
luminosity evaluation. The numbers of events were then corrected for
acceptance and data acquisition efficiency \emph{etc.} and presented
in $0.25^{\circ}$ bins of deuteron laboratory angle in
Fig.~\ref{Wig05}.

%
%
%
%

Isospin invariance requires that the differential cross section for
$np\to d\pi^0$ should be half that of the $pp\to d\pi^+$ reaction,
for which there are many measurements and an extensive data
compilation by the SAID group~\cite{SAIDpid}. Predictions of the SAID
program reproduce well the shape of the data in Fig.~\ref{Wig05} and,
after scaling this to agree with our experimental points, the
luminosity can be deduced. It is of course possible that there could
be small isospin violations between $\pi^0$ and $\pi^+$ production
which may introduce uncertainties in the luminosity on the very few
per cent level.

The luminosity determined in this way corresponds to that of the
unpolarised mode. However, the orbit of the deuterons inside COSY
should be independent of the polarisation mode of the ion source and so
the relative luminosity for the other modes can be evaluated using
the information provided by the beam-current transformer
(BCT)~\cite{Chiladze1}.

%
%
\subsection{Deuteron--proton elastic scattering}
\label{dpelastic} \setcounter{equation}{0}

\begin{figure}[htb]
\begin{center}
\resizebox{0.44\textwidth}{!}{\includegraphics{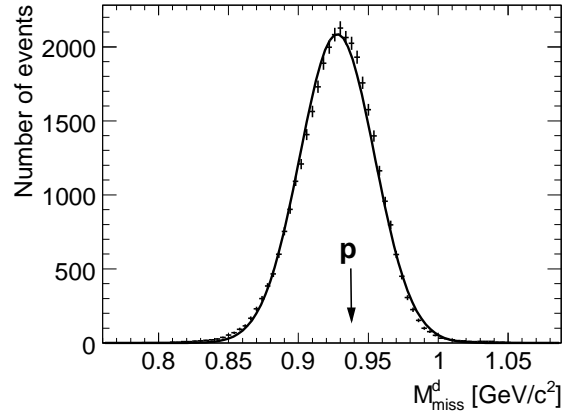}}
\caption{The missing mass with respect of the observed deuterons for
events that are close to the expected $dp$ elastic locus in
Fig.~\ref{Wig02}. There is little or no background under the proton
peak, which has a width of $\sigma=27$~MeV/c$^2$.} \label{Wig06}
\end{center}
\end{figure}

An alternative way of determining the luminosity required to evaluate
the charge--exchange cross section would be through the measurement
of deuteron--proton elastic scattering using data from the
unpolarised spin mode. Due to its very high cross section, the fast
deuterons from this process are clearly seen in the angle--momentum
plot of Fig.~\ref{Wig02} for laboratory polar angles from $5^{\circ}$
to $10^{\circ}$. Since the locus of this reaction is well separated
from those of the others, it is to be expected that the background
should be very small. The justification for this is to be found in
Fig.~\ref{Wig06} where, after selecting events from a broad region
around the $(p,\theta_{xz})$ locus, the missing mass with respect to
the deuteron shows a proton peak with negligible background. As
discussed below, the very different populations along the locus is
merely a reflection of the rapid variation of the differential cross
section with angle.
%
%
%
%

\begin{figure}[htb]
\begin{center}
\resizebox{0.44\textwidth}{!}{\includegraphics{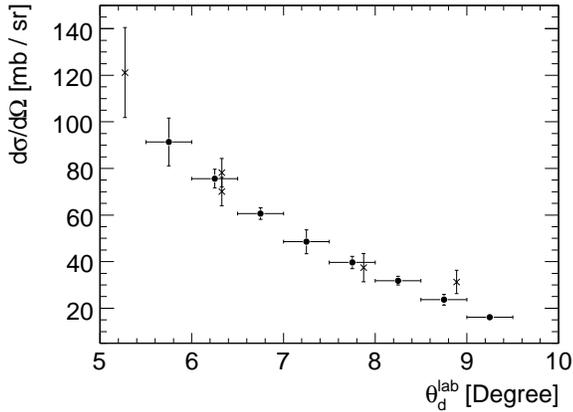}}
\caption{Laboratory differential cross section for small angle $dp$
elastic scattering in $0.5^{\circ}$ bins (circles) are compared to
the $pd\to pd$ values reported in Ref.~\cite{Boschitz} and
transformed to the proton rest frame (crosses). By scaling our data
to agree with these values, an estimate of the luminosity could be
made. Only data in the range $5.5^{\circ}<\theta_{d}^{\rm lab} <9.5^{\circ}$
were used for this purpose.\label{Wig07}}
\end{center}
\end{figure}

Having identified good $dp$ elastic scattering events, their
numbers were corrected for the MWPC efficiency. For this purpose,
two--dimensional efficiency maps were created for each plane and the
tracks weighted using these maps. The events were grouped into
laboratory polar angular bins of width 0.5$^\circ$ in order to plot
the angular distribution. The numbers in each bin were adjusted by
the prescaling factor using the correction of the DAQ efficiency. The
resulting differential cross section is plotted as a function of the
deuteron laboratory angle in Fig.~\ref{Wig07}, using the
normalisation discussed below.
%
%
%
%

Very close to our energy ($T_d/2 = 585$~MeV) elastic proton--deuteron
scattering was measured at 582~MeV using carbon and deuterated
polyethylene targets together with counter
telescopes~\cite{Boschitz}. The differential cross sections were then
obtained from a CD$_2$--C subtraction. The resulting values,
transformed to the proton rest frame, are also shown in
Fig.~\ref{Wig07}. Although the absolute normalisation was established
well using the carbon activation technique, it should be noted that a
test measurement at one angle, where a magnetic spectrometer was used
to suppress background from breakup protons, led to a cross section
that was 10\% lower, though with a significant statistical error.
There is therefore the possibility that these data include some
contamination from non-elastic events. Despite this uncertainty, the
comparison of the two data sets allows a value of the luminosity to
be deduced for our experiment. To avoid regions of strong azimuthal
variation in the ANKE acceptance, only the range
$5.5^{\circ}<\theta_{d}^{\rm lab} <9.5^{\circ}$ was considered for this
evaluation.

The only other available data close to our momentum ($p_d=
2.4$~GeV/c) come from a measurement of deuteron--proton elastic
scattering in a hydrogen bubble chamber experiment at ten momenta
between 2.0 and 3.7~GeV/c~\cite{Katayama}. Although numerical values
are not available, the results show a smooth variation with beamRe: Wednesday still
momentum when plotted as a function of the momentum transfer $t$.
Interpolating these results to 582~MeV, the data seem to be
consistent with those of Ref.~\cite{Boschitz}, though the variation
of the cross section with $t$ is extremely strong.

\begin{figure}[hbt]
\begin{center}
\resizebox{0.44\textwidth}{!}{\includegraphics{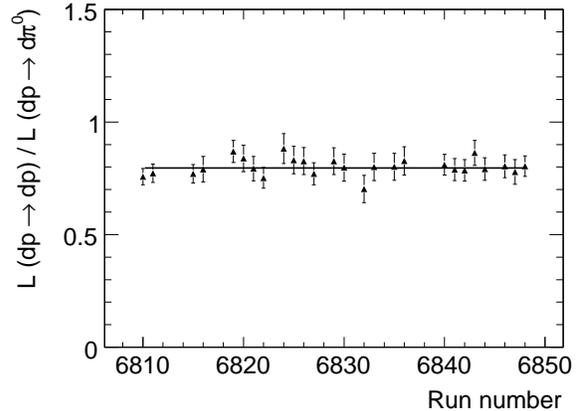}}
\caption{Ratio of the luminosity determined from small angle
deuteron--proton elastic scattering and quasi--free $np\to d\pi^0$
pion production \emph{versus} the individual run number.}
\label{Wig08}
\end{center}
\end{figure}

\subsection{Comparison of the luminosity determinations}
\label{comparison} \setcounter{equation}{0}

Having determined the luminosity independently on the basis of the
$dp\to dp$ and quasi--free $np\to d\pi^0$ measurements, the results
are compared in Fig.~\ref{Wig08} for all the individual ``good''
runs. The luminosity ratio is consistent with being constant,
\begin{equation}
\left.\mathcal{L}(dp\to dp)\right/\mathcal{L}(np\to d\pi^0) =
0.80\pm0.01\,
\end{equation}
where the error is purely statistical. The smallness of the
fluctuations in Fig.~\ref{Wig08} implies that the two methods are
sensitive to the same quantity, though with a different overall
normalisation. Of the 20\% discrepancy, about 5\% can be accounted
for by the shadowing correction~\cite{Glauber}, which reduces
slightly the quasi--free cross sections on the deuteron compared to
their free values. To a first approximation the deuteron charge
exchange would be subject to a rather similar shadowing correction.
Some of the residual difference might be due to inelastic events in
the published data~\cite{Boschitz}.

Apart from the shortages in the World data set on $dp\to dp$ compared
to $pp\to d\pi^+$, it should be noted that the elastic
deuteron--proton differential cross section varies very rapidly with
angle. A shift of a mere $0.1^{\circ}$ in the deuteron laboratory
angle induces a 5\% change in the cross section. This is to be compared to the
absolute precision in the angle determination in ANKE, which is
$\approx 0.2^{\circ}$. For these reasons
much more confidence can be ascribed to the quasi-free $np\to d\pi^0$
method to determine the luminosity, which we believe to be accurate
to better than about 4\%, based upon the study of the errors quoted in
$pp\to \pi^+d$  measurements in this energy region~\cite{SAIDpid}. The
resulting integrated luminosity for the unpolarised mode was
$\mathcal{L}=(12.5\pm0.5)$~nb$^{-1}$.

%
%
\section{Deuteron charge--exchange cross section}
\label{dce}\setcounter{equation}{0}

\begin{figure}[htb]
\begin{center}
\resizebox{0.44\textwidth}{!}{\includegraphics{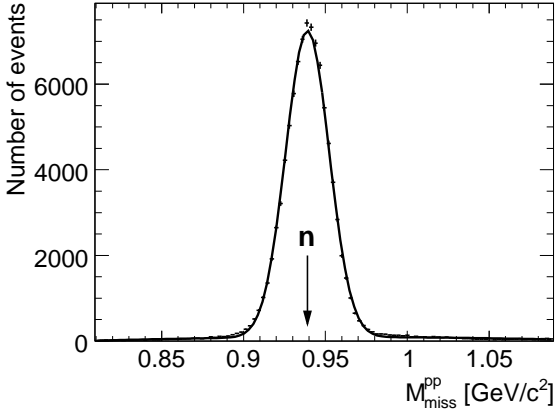}}
\caption{Missing mass distribution for proton pairs selected by the
TOF criterion described in the text. A fit to the data in terms of a
Gaussian plus a smoothly varying background shows the latter to be at
about the 2\% level. The central value agrees with the neutron mass
to within 0.1\%. Events falling within $\pm2.5\sigma$ of the peak
position were retained in the analysis.} \label{Wig09}
\end{center}
\end{figure}

The deuteron charge exchange on hydrogen, $dp\to \{pp\}n$ is defined
to be the reaction where the diproton emerges with low excitation
energy $E_{pp}$. When this takes place with small momentum transfer,
the two fast protons are emitted in a narrow forward cone with
momenta around half that of the deuteron beam. As described fully in
Ref.~\cite{Chiladze0}, such coincident pairs can be clearly
identified using information from the FD system in much the same way
as for the $dp\to p_{\rm sp}d\pi^0$ reaction of Sec.~\ref{pizero}.
Having measured the momenta of two charged particles, their times of
flight from the target to the hodoscope were calculated assuming that
these particles were indeed protons. The difference between these two
times of flight was compared with the measured time difference for
those events where the particles hit different counters in the
hodoscope. This selection, which discarded about 20\% of the events,
eliminated almost all the physics background, for example, from $dp$
pairs associated with $\pi^0$ production. The resulting missing--mass
distribution for identified $ppX$ events shows a clean neutron peak
in Fig.~\ref{Wig09} at $M_X=(940.4\pm0.2)$~MeV/c$^2$ with a width of
$\sigma = 13$~MeV/c$^2$, sitting on a slowly varying 2\% background.

\begin{figure}[htb]
\begin{center}
\resizebox{0.44\textwidth}{!}{\includegraphics{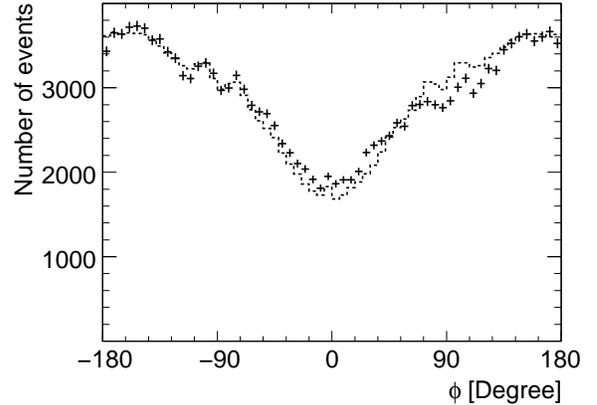}}
\caption{Distribution of $dp\to \{pp\}n$ events in the azimuthal
angle $\phi$ obtained with an unpolarised beam for $E_{pp}<3$~MeV
(dashed) compared to a simulation of expected events (crosses). }
\label{Wig10}
\end{center}
\end{figure}

Only at small momentum transfer and small $pp$ excitation energy is
the ANKE geometric acceptance even approximately isotropic. Unlike
the case of $dp\to p_{\rm sp}d\pi^0$ used for the luminosity
determination, one clearly cannot limit the data selection to this
small region of phase space. Figure~\ref{Wig10} shows the
distribution of unpolarised charge--exchange events for $E_{pp}<
3$~MeV in terms of the azimuthal angle of the diproton $\phi$ measured with
respect to the COSY plane. 
This variable is of critical importance in the
separation of the deuteron analysing powers for the \dpcepol\
reaction and so it is necessary to have a reasonable understanding of
its behaviour within a reliable GEANT simulation. As can be seen from
Fig.~\ref{Wig10}, this has been successfully achieved.

\begin{figure}[htb]
\begin{center}
\resizebox{0.44\textwidth}{!}{\includegraphics{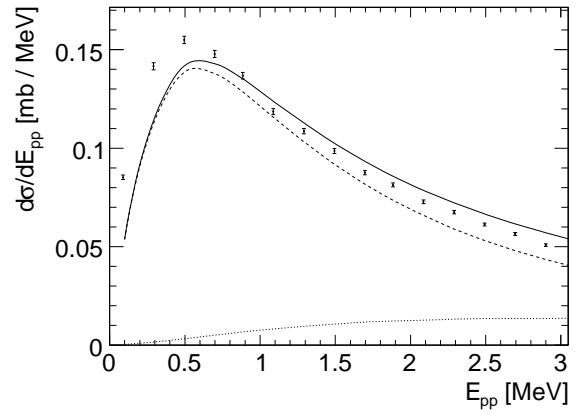}}
\caption{Differential cross section for unpolarised $dp\to \{pp\}n$
integrated over momentum transfer $q<100$~MeV/c as a function of the
excitation energy $E_{pp}$. Only statistical errors are shown.
The impulse approximation predictions are
shown separately for the $^1\!S_0$ (dashed) and higher waves
(dot-dashed) as well as their sum (solid curve).} \label{Wig11}
\end{center}
\end{figure}

Since the counting rate varies rapidly with both $E_{pp}$ and $q$,
the acceptance was estimated by inserting the predictions of the
impulse approximation model into the Monte Carlo simulation in a
two--dimensional grid. Having then corrected the numbers of events
for acceptance and DAQ and other efficiencies, the cross sections
found on the basis of the quasi-free $np\to d\pi^0$ luminosity were
put in ($E_{pp},q$) bins. The results obtained by summing these data
over the interval in momentum transfer $0<q<100$~MeV/c are presented
as a function of $E_{pp}$ in Fig.~\ref{Wig11}.

The impulse approximation predictions, also shown in
Fig.~\ref{Wig11}, describe these data reasonably well even in
absolute magnitude, although the model seems to be pushed to slightly
higher values of $E_{pp}$ than the data. It is important to note
that, even for excitation energies as low as 3~MeV, there are
significant contributions from higher partial waves. These arise
preferentially for this reaction because even a small momentum kick
to the neutron when it undergoes a charge exchange can induce high
partial waves because of the large deuteron radius. This is to be contrasted
with large momentum transfer deuteron break-up where the interaction is of
much shorter range~\cite{Komarov}.

\begin{figure}[htb]
\begin{center}
\resizebox{0.44\textwidth}{!}{\includegraphics{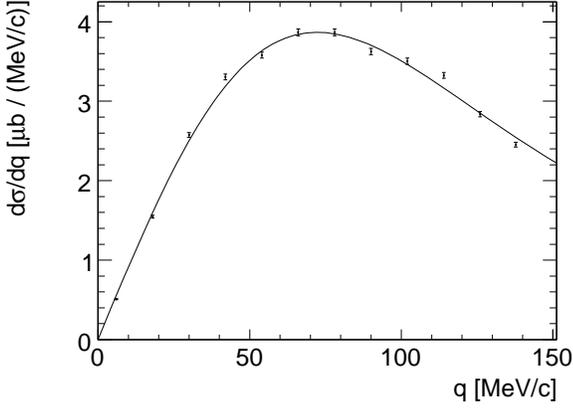}}
\caption{Unpolarised differential cross section for the \dpce\
reaction for $E_{pp}<3$~MeV compared with the impulse approximation
predictions. Only statistical errors are shown. There is in addition a global
systematic uncertainty of about 6\%.} \label{Wig12}
\end{center}
\end{figure}

The variation of the cross section with momentum transfer can be
found in Fig.~\ref{Wig12} for $E_{pp}<3$~MeV. The impulse
approximation of section~\ref{Impulse} also describes well the
dependence on this variable out to $q=140$~MeV/c. Once again it
should be noted that no adjustment has been made to the model or the
experimental data; the luminosity was evaluated independently using
the quasi-free $np\to d\pi^0$ reaction.

%
%
\section{Deuteron beam polarisation}
\label{polarisation}\setcounter{equation}{0}

The COSY polarised ion source that feeds the circulating beam was
programmed to provide a sequence of one unpolarised state, followed
by seven combinations of deuteron vector ($P_z$) and tensor
($P_{zz}$) polarisations, where $z$ is the quantisation axis in the
source frame of reference. The beam polarisation was measured in the
COSY ring during the acceleration at a beam energy of $T_d=270$~MeV
using the EDDA polarimeter~\cite{EDDA}. The measurement of a variety
of nuclear reactions in ANKE did not show any loss of
beam polarisation when the deuterons were subsequently brought up to
the experimental energy of $T_d=1170$~MeV~\cite{Chiladze1}.

\begin{table*}[htb]
\begin{small}
\begin{center}
\begin{tabular}{|c|c|c||c|c|c|}
\hline Mode--n$\vphantom{\Big[}$& $P_z^{{\rm Ideal}}$ &$P_z^{{\rm
EDDA}}$ & $P_{zz}^{{\rm Ideal}}$ &
$P_{zz}^{{\rm EDDA}}$ & $P_{zz}^{{\rm Sta}}$\\
\hline
$0$ &0    & 0                         &\phantom{-}0&0                           & $-0.006\pm 0.016$\\
$1$ &--2/3&$-0.499\pm0.021$           &\phantom{-}0& $\phantom{-}0.057\pm0.051$ & $\phantom{-}0.040\pm0.016$\\
$2$ &+1/3 &$\phantom{-}0.290\pm0.023$ &+1          & $\phantom{-}0.594\pm0.050$ & $\phantom{-}0.658\pm0.032$\\
$3$ &--1/3&$-0.248\pm0.021$           &--1         & $-0.634\pm0.051$           & $-0.575\pm0.032$\\
$4$ &+1/2 &$\phantom{-}0.381\pm0.027$ &$-1/2$     & $-0.282\pm0.064$           & $-0.359\pm0.024$\\
$5$ &--1  &$-0.682\pm0.027$           &+1          & $\phantom{-}0.537\pm0.064$ & $\phantom{-}0.594\pm0.030$\\
$6$ &+1   &$\phantom{-}0.764\pm0.027$ &+1          & $\phantom{-}0.545\pm0.061$ & $\phantom{-}0.440\pm0.024$\\
$7\vphantom{\int}$ &--1/2&$-0.349\pm0.027$           &$-1/2$     & $-0.404\pm0.065$           & $-0.355\pm0.024$\\
\hline
\end{tabular}
\end{center}
\end{small}
\caption{The configurations of the polarised deuteron ion source,
showing the ideal values of the vector and tensor polarisations and
their measurement using the EDDA polarimeter at a beam energy of
$T_d=270$~MeV~\cite{EDDA}. The standardised values of $P_{zz}$
obtained on the basis of all the deuteron charge--exchange data are
given in the final column. However, it should be noted that mode--0
was indeed completely unpolarised and the statistical error quoted
here is merely to show that the charge--exchange data were completely
consistent with that.\label{table1}}
\end{table*}

\begin{figure}[htb]
\begin{center}
\resizebox{0.40\textwidth}{!}{\includegraphics{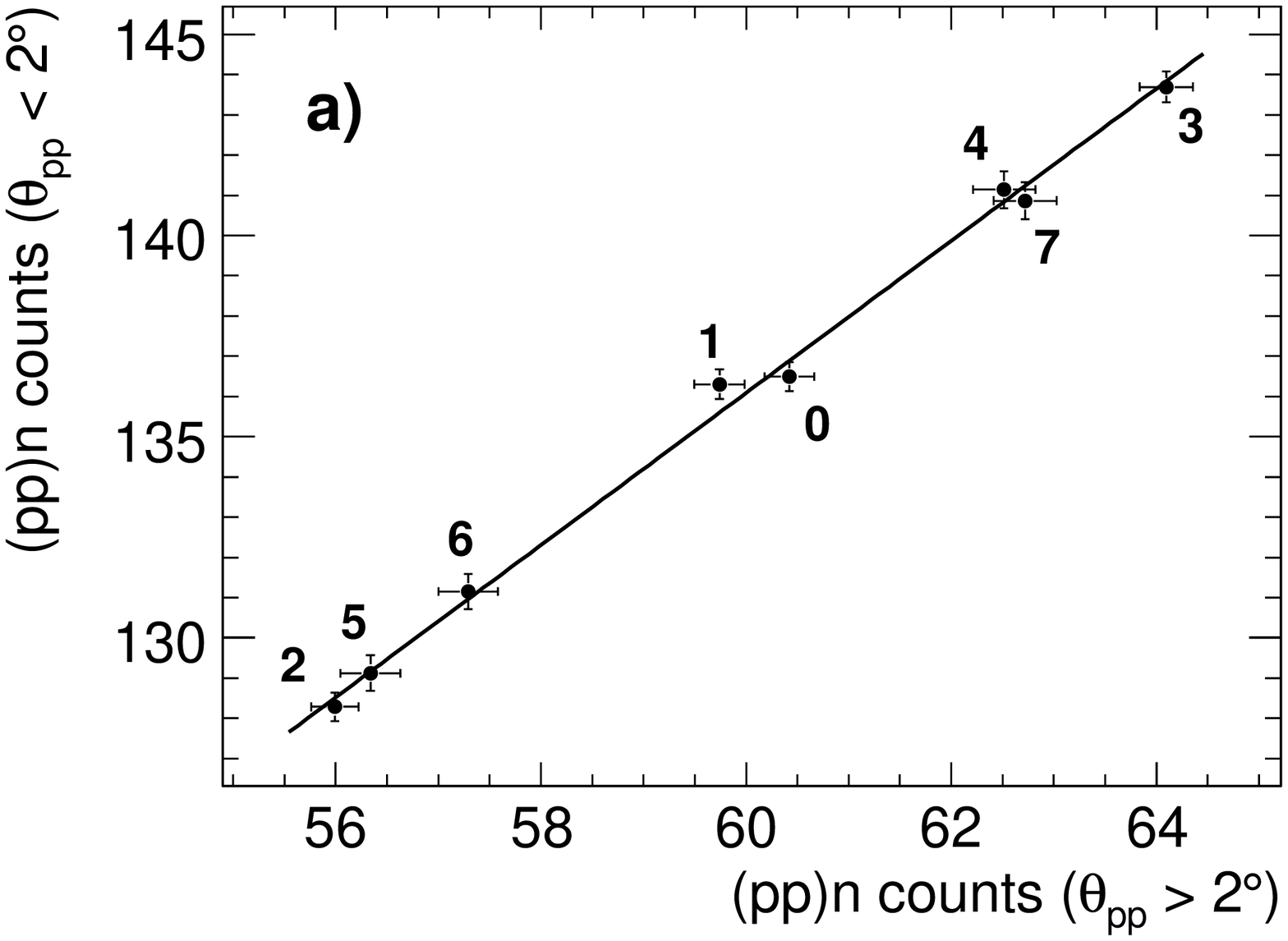}}
\resizebox{0.40\textwidth}{!}{\includegraphics{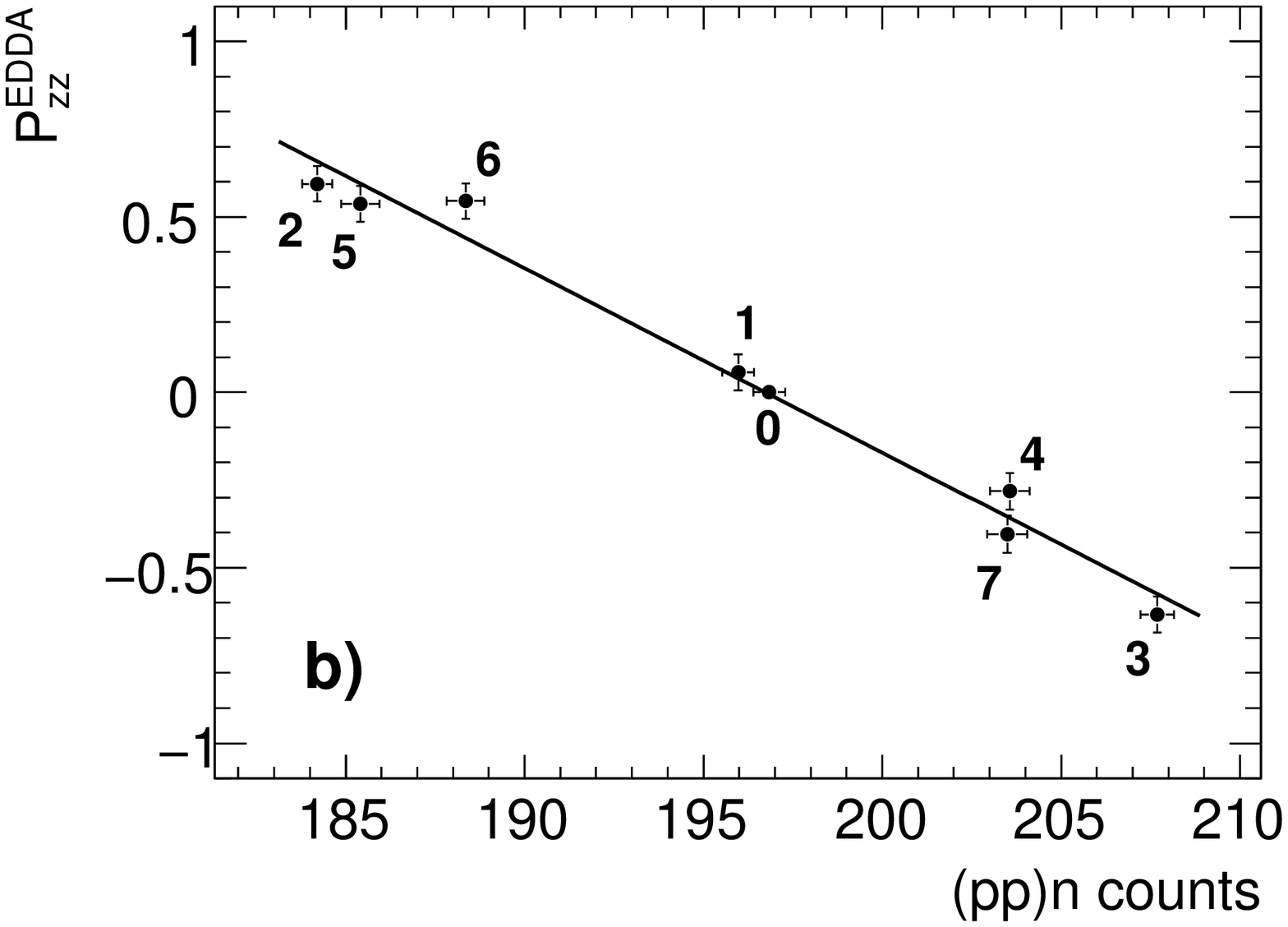}}
\caption{Normalised counts $\times 10^{-3}$ for the \dpce\ reaction for the eight
different source modes of Table~\ref{table1}, (a) for events where
the diproton laboratory angle is less than $2^{\circ}$ compared to events where
the angle is greater than $2^{\circ}$, and (b) compared to the EDDA
measurements of the beam tensor polarisation~\cite{EDDA}. Also shown
are straight line fits to the data.} \label{Wig13}
\end{center}
\end{figure}

Although the EDDA systematic uncertainties are quite low, as can be
seen from Table~\ref{table1}, only limited statistics were collected
and this we alleviate by using the internal consistency of our
deuteron charge--exchange data themselves. According to impulse
approximation predictions~\cite{BW,Carbonell}, the deuteron vector
analysing power for the \dpcepol\ reaction should vanish for small
excitation energies. Since the values of $P_z$ can have no effect for
$\theta_{pp}\approx 0$, this hypothesis can be tested by comparing
the charge--exchange count rates, normalised on the BCT, for small
and larger diproton angles. Any deviations from linearity could be
ascribed to a $it_{11}$ dependence since Table~\ref{table1} shows
that the eight modes have widely different values of $P_z$. All the
data presented in Fig.~\ref{Wig13}a fit well to a straight line,
which reinforces the belief that the charge exchange is, as expected,
only sensitive to the value of $P_{zz}$.

In Fig.~\ref{Wig13}b the totality of the charge--exchange data is
compared to the values of $P_{zz}$ measured with the EDDA
polarimeter. The scatter is larger due to the EDDA statistical errors
but a linear fit is a good representation of the data. We then
replace the EDDA values of $P_{zz}$ for each of the individual modes
by those corresponding to the linear regression shown in
Fig.~\ref{Wig13}b and these standardised values are given in
Table~\ref{table1}. This procedure retains the average dependence on
the EDDA polarisations while reducing the statistical fluctuations
inherent therein.

Although in the earlier work~\cite{Chiladze1} the results were
presented in terms of Cartesian analysing powers, the extrapolation
to $q=0$ is more stable when linear combinations corresponding to
those in a spherical basis are used. The relation between the two
is~\cite{Ohlsen}
\begin{eqnarray}
\nonumber A_{yy}&=& -t_{20}/\sqrt{2} -\sqrt{3}\,t_{22}\,,\\
\label{carttopol} A_{xx}&=& -t_{20}/\sqrt{2} +\sqrt{3}\,t_{22}\,.
\end{eqnarray}

The differential cross section for a polarised \dpcepol\ reaction
then becomes
\begin{eqnarray}\nonumber
\left.\frac{\dd\sigma}{\dd t}(q,\phi)\right/
\left(\frac{\dd\sigma}{\dd t}(q)\right)_{\!\!0}=
1+\sqrt{3}P_{z}it_{11}(q)\cos\phi\\
-\frac{1}{2\sqrt{2}}P_{zz}t_{20}(q)
-\frac{\sqrt{3}}{2}P_{zz}t_{22}(q)\cos(2\phi)\,,
\label{polarisation2}
\end{eqnarray}
where the $0$ subscript refers to the unpolarised cross section. 
We are here using a right-handed coordinate system where the beam defines the
$z$--direction and $y$ is along the upward normal to the COSY orbit. The polar
angle $\theta$ is measured with respect to the $z$--axis and the azimuthat
with respect to the $x$.
%
%
\section{Analysing powers of the deuteron charge--exchange reaction}
\label{Ap}\setcounter{equation}{0}

\begin{figure}[htb]
\begin{center}
\resizebox{0.40\textwidth}{!}{\includegraphics{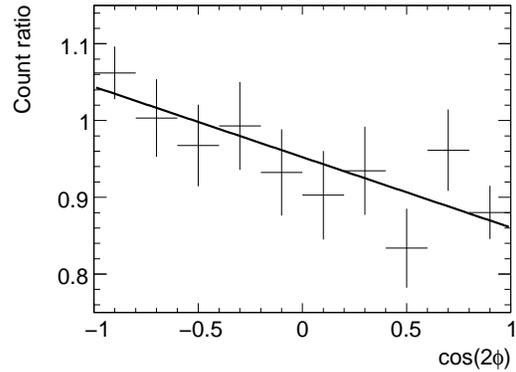}}
\caption{Azimuthal angular dependence of the ratio of the normalised
\dpce\ count rates for the source mode--5 to the average of
modes--0 and --1, which have no tensor polarisation. Here the data are
shown for the bin $40 < q < 60$~MeV/c and $E_{pp}<1$~MeV. The linear
fit in $\cos 2\phi$ allows the analysing power to be extracted by
fitting the data to the right hand side of Eq.~(\ref{polarisation2}).
} \label{Wig14}
\end{center}
\end{figure}

Having identified the charge--exchange events, as described for the
unpolarised case of section~\ref{dce}, the data were corrected for
beam current and dead time and placed in 20~MeV/c bins in
$q$ and ten in $\cos2\phi$. This procedure was carried out for two ranges in
excitation energy, $0.1<E_{pp}<1$~MeV and $1<E{pp}<3$~MeV. Although it is
clear from Fig.~\ref{Wig10} that the acceptance in terms of the
azimuthal angle $\phi$ is well reproduced by the simulation, we have
used modes--0 and --1, where there is zero tensor polarisation, to
provide the best estimate of the denominator in
Eq.~(\ref{polarisation2}). By doing this we are using the fact that
the geometric acceptances should be universal, \emph{i.e.},
independent of the polarisation mode of the ion source. An example of the
linear fit is shown in Fig.~\ref{Wig14} for polarisation mode--5.

The analysing powers of the \dpcepol\ reaction were subsequently
evaluated by fitting with Eq.~(\ref{polarisation2}) and using 
the beam polarisations of modes--2 to --7
quoted in Table~\ref{table1}. An estimate of the statistical errors
inherent in this procedure could be obtained by studying the scatter
of the results for these six polarisation modes of the source. A
similar procedure in terms of $\cos\phi$ allowed bounds to be
obtained on the vector analysing power but, as expected from both
theory and the linearity of Fig.~\ref{Wig13}a, all the data are
consistent with $it_{11}$ vanishing within error bars. The averages
over the whole $q$ range are $<it_{11}> = -0.001\pm 0.004$ for
$E_{pp}<1$~MeV and $-0.004\pm 0.004$ for $1<E_{pp}<3$~MeV.

\begin{figure}[htb]
\begin{center}
\resizebox{0.40\textwidth}{!}{\includegraphics{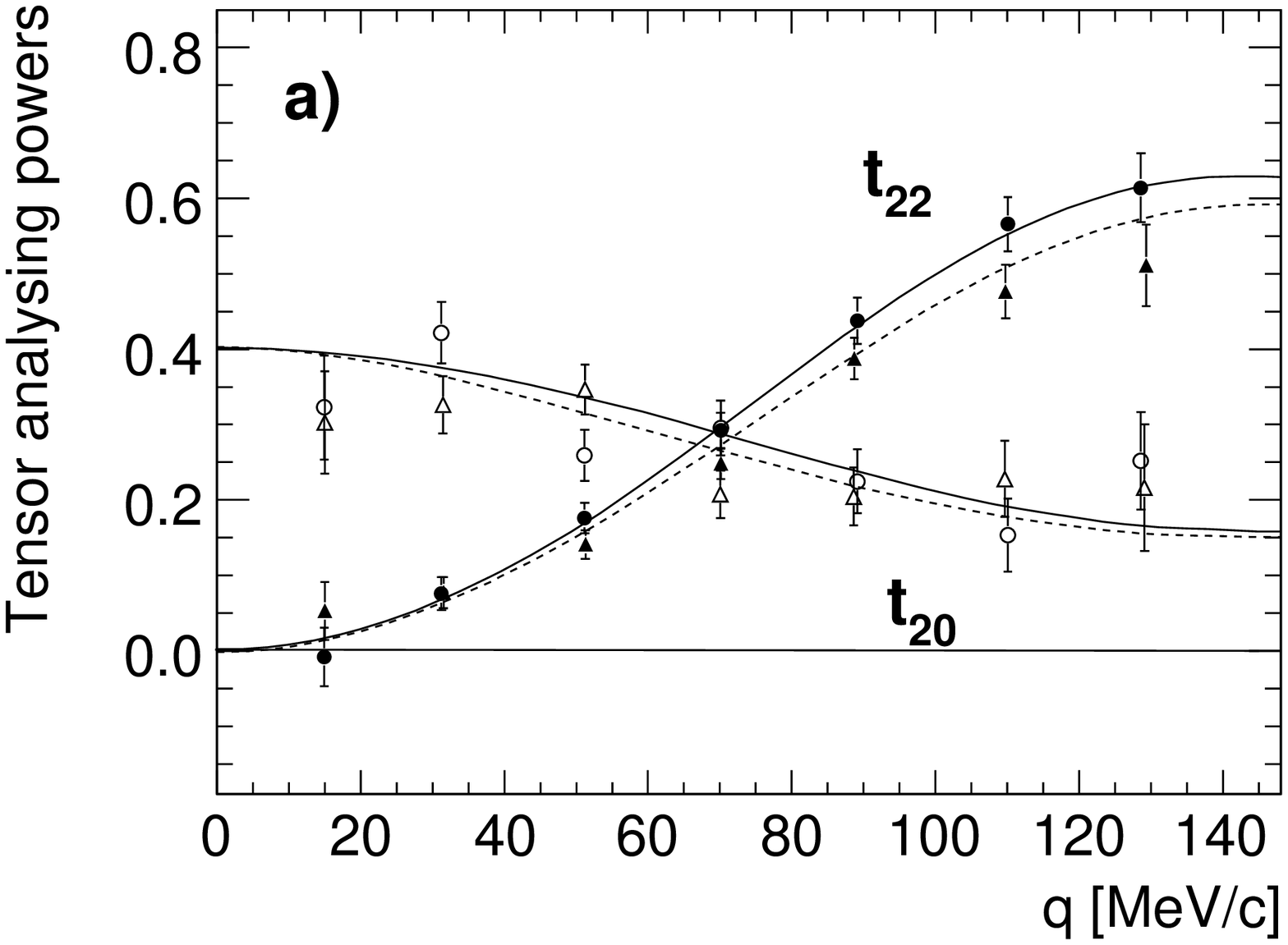}}
\resizebox{0.40\textwidth}{!}{\includegraphics{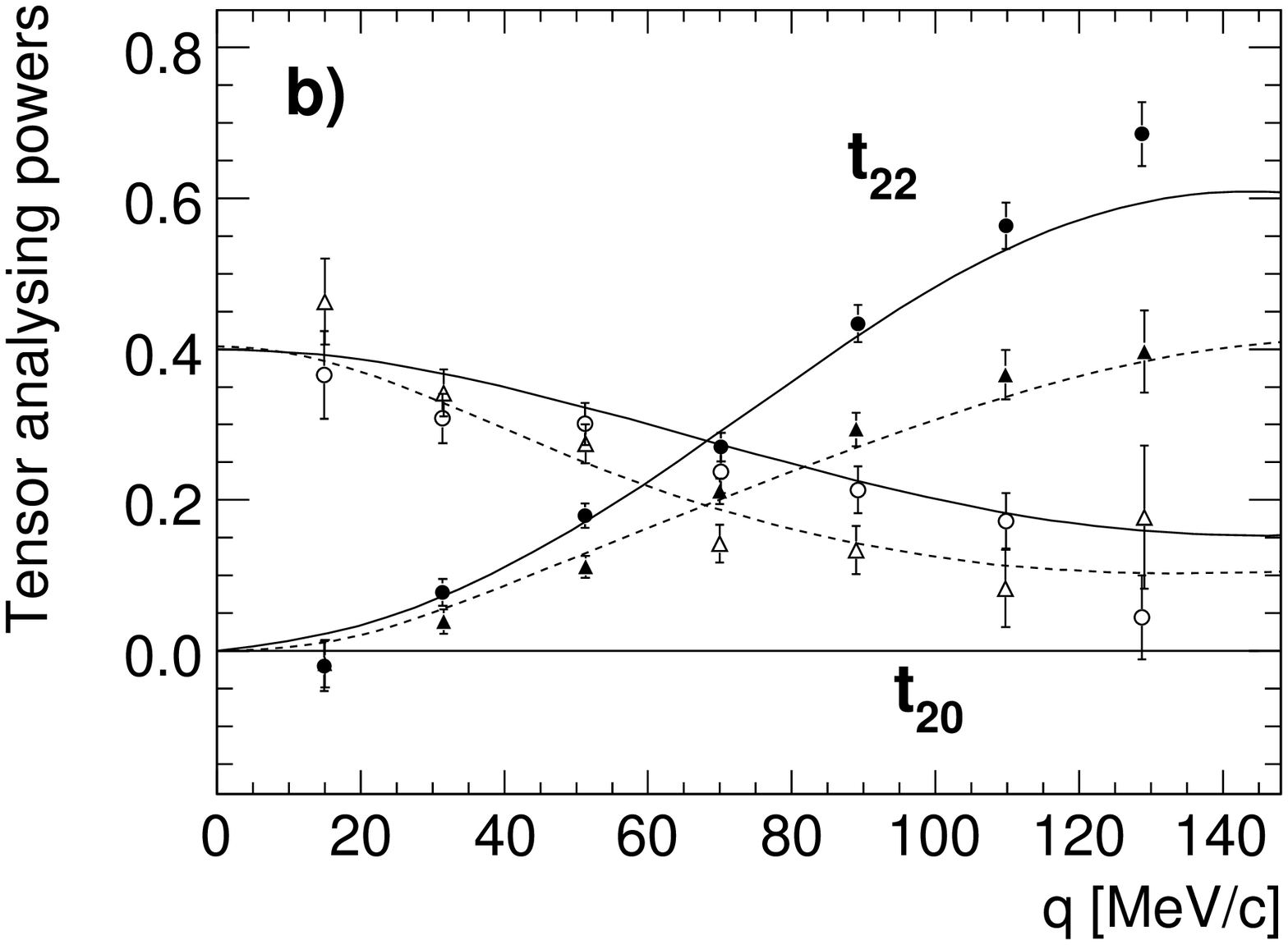}}
\caption{Spherical tensor analysing powers $t_{20}$ (open symbols)
and $t_{22}$ (closed symbols) for the \dpce\ reaction at
$T_d=1170$~MeV for (a) $0.1 < E_{pp} < 1.0$~MeV and (b)$1.0 < E_{pp}
< 3.0$~MeV. The circles correspond to data where $|\cos\theta_{qk}| <
0.5$ while the triangles denote the results for $|\cos\theta_{qk}| >
0.5$. The solid and dashed curves are the impulse approximation
predictions for the same angular selections, respectively.}
\label{Wig15}
\end{center}
\end{figure}

Our experimental values of the two tensor analysing powers are shown
in Fig.~\ref{Wig15} for the two ranges in $E_{pp}$ as a function of
the momentum transfer. The signals both fall when $E_{pp}$ rises due
to the influence of higher partial waves. This dilution can be
partially offset by making a cut on the angle between $\vec{q}$ and
$\vec{k}$ since $P$--waves are not excited when
$\vec{q}\cdot\vec{k}=0$~\cite{BW}. Therefore the data with small
values of $|\cos\theta_{qk}|$ are far less affected by the $E_{pp}$
cut.

The rapid rise of $t_{22}$ with $q$ is mainly a result of the fall in
the $\delta(q)$ amplitude which, in a simple absorbed
one--pion--exchange model, should vanish for $q\approx m_{\pi}c$. The
behaviour can therefore be understood semi--quantitatively on the
basis of Eq.~(\ref{impulse}). The much smoother variation of $t_{20}$
is also expected, with a gentle decline from the forward value, once
again being mainly driven by the fall in the $\delta(q)$ amplitude.
All these features are well reproduced by the impulse approximation
model~\cite{Carbonell} using reliable $np$ amplitudes~\cite{SAIDNN}.

Although all the experimental data agree with the impulse
approximation model one could, of course, invert the question. How
well could one determine the amplitudes if there were no information
available from the $np$ phase shifts? Although the data reported here
were obtained over only two--day run, these are already sufficient to
determine quite well the ratio of the $|\varepsilon(0)|/|\beta(0)|$
in the forward direction. Since little dilution of the $t_{20}$
signal is expected at $q=0$, all the data for $E_{pp}<3$~MeV were
fitted to a quadratic in $q^2$ for $q\leq 100$~MeV/c. The value
obtained at the origin gives $t_{20}= 0.37\pm0.02$, where the error
is purely statistical. The uncertainty introduced by the beam
polarisation would, however, contribute less than $\pm0.01$ to this.
Since there is little or no dilution of the analysing power by the
$P$-waves at $q=0$, using Eq.~(\ref{impulse}), this result translates 
into an amplitude ratio of
\begin{equation}
|\varepsilon(0)|/|\beta(0)| = 0.61 \pm 0.03\,.
\end{equation}

The precision here is, of course, better than that which is achieved
for the absolute value of the forward amplitudes, 
where the overall normalisation and other effects
introduce another 3\% uncertainty.

%
%
\section{Conclusions}
\label{Conclusions}

In this pilot study we have shown that the measurement of the
differential cross section and two deuteron tensor analysing powers
of the \dpcepol\ reaction at 585~MeV per nucleon allows one to deduce
values of the magnitudes of the amplitudes $|\beta(q)|^2+|\gamma(q)|^2$,
$|\delta(q)|^2$, and $|\varepsilon(q)|^2$. The results achieved agree
very well with modern phase shift analyses~\cite{SAIDNN}. There is no
obvious reason why this success should not be repeated at higher
energies where the neutron--proton database has far more ambiguities.

In addition to extending the ANKE measurements to the maximum COSY
energy of 1.15~GeV per nucleon, experiments are being undertaken with
polarised beam and target~\cite{Gomez}. The values of the two vector
spin--correlation parameters depend upon the interferences of
$\varepsilon$ with the $\beta$ and $\delta$
amplitudes~\cite{Barbaro}. Furthermore, the use of inverse kinematics
with a polarised proton incident on a polarised deuterium gas cell~\cite{PIT}
will allow the study to be continued up to 2.9~GeV per
nucleon~\cite{SPIN}. In future experiments an independent check on
the luminosity will be provided through the study of the energy loss
of the circulating beam in COSY~\cite{Stein}.

On the other hand the low excitation energy charge exchange on the
deuteron gives no direct information on the spin--independent
amplitude $\alpha$, whose magnitude can only be estimated by
comparing the deuteron data with the free $np\to pn$ differential
cross section. It is seen, for example, from Eq.~(\ref{Dean}), that
the value of $|\alpha(0)|^2$ can be
determined with respect to the other amplitudes by measuring the
ratio of the charge exchange on the deuteron and proton~\cite{Dean}.

At $q=0$ there is potential redundancy between the measurement of the
\dpce\ and $np\to pn$ cross sections, though the normalisation is
much easier to achieve with a beam of charged particles. Using this
information in association with data on total cross section
differences, it seems likely that a clear picture of the
neutron--proton charge--exchange amplitudes in the forward direction
will emerge~\cite{Lehar}.

%
%
\begin{acknowledgement}
We are grateful to R.~Gebel, B.~Lorentz, H.~Rohdje\ss, and D.~Prasuhn
and other members of the accelerator crew for the reliable operation
of COSY and the deuteron polarimeters. We would like to thank
I.I.~Strakovsky for providing us with up--to--date neutron--proton
amplitudes. We have also profited from discussions with F.~Lehar.
This work has been supported by the COSY FFE program, HGF--VIQCD, and
the Georgian National Science Foundation Grant (GNSF/ST06/4-108).
\end{acknowledgement}

%
%


\begin{thebibliography}{99}
%
\bibitem{SAIDNN} R.A.~Arndt, I.I.~Strakovsky, R.L.~Workman,
Phys.\ Rev.\ C \textbf{62}, 034005 (2000);
\verb=http://gwdac.phys.gwu.edu=.
%
\bibitem{Lehar} F.~Lehar, C.~Wilkin, Eur.\ Phys.\ J.\ A
\textbf{37}, 143 (2008).
%
\bibitem{Dean}  N.W.~Dean, Phys.\ Rev.\ D \textbf{5}, 1661 (1972);
                N.W.~Dean, Phys.\ Rev.\ D \textbf{5}, 2832 (1972).
%
\bibitem{Sharov} V.I.~Sharov \emph{et al.}, Czech.\ J.\ Phys.\
\textbf{56}, F117 (2006); \emph{idem} Dubna preprint E1-2008-61
(2008).
%
\bibitem{BW} D.V.~Bugg, C.~Wilkin, Nucl.\ Phys.\ A \textbf{467}, 575 (1987).
%
\bibitem{Ellegaard} C.~Ellegaard \emph{et al.}, Phys.\ Rev.\ Lett.\ \textbf{59}, 974 (1987).
%
\bibitem{Kox} S.~Kox \emph{et al.}, Nucl.\ Phys.\ A \textbf{556}, 621 (1993).
%
\bibitem{SPIN} A.~Kacharava, F.~Rathmann, C.~Wilkin, \emph{Spin Physics
from COSY to FAIR}, COSY proposal \textbf{152} (2005), arXiv:nucl-ex/0511028.
%
\bibitem{PIT} K.~Grigoryev \emph{et al.}, AIP Conf.\ Proc.\  \textbf{915}, 979 (2007).
%
\bibitem{Chiladze0} D.~Chiladze \emph{et al.}, Phys.\ Lett.\ B \textbf{637}, 170 (2006).
%
\bibitem{Carbonell} J.~Carbonell, M.B.~Barbaro, C.~Wilkin,
Nucl.\ Phys.\ A \textbf{529}, 653 (1991).
%
\bibitem{Chiladze1} D.~Chiladze \emph{et al.}, Phys.\ Rev.\ ST Accel.\ Beams
\textbf{9}, 050101 (2006).
%
\bibitem{Mai97} R.~Maier \emph{et al.}, Nucl.\ Instrum.\ Methods  A
  \textbf{390}, 1 (1997).
%
\bibitem{ANKE}
S.~Barsov \emph{et al.}, Nucl.\ Instr.\ Meth. A \ \textbf{462}, 364
(2001).
%
\bibitem{Dymov} S.~Dymov \emph{et al.}, Part.\ Nucl.\ Lett.\
\textbf{1}, 40 (2004).
%
\bibitem{Khoukaz}
A.~Khoukaz \emph{et al.}, Eur.\ Phys.\ J.\ D \textbf{5}, 275 (1999).
%
\bibitem{PLUTO} Pluto WEB page:
\verb=http://www-hades.gsi.de/computing/=\\
\verb=pluto/html/PlutoIndex.html=.
%
\newpage
%
\bibitem{Paris} M.~Lacombe \emph{et al.}, Phys.\ Lett.\
\textbf{101B}, 139 (1981).
%
\bibitem{SAIDpid} R.A.~Arndt, I.I.~Strakovsky, R.L.Workman, D.V.~Bugg,
Phys.\ Rev.\ C \textbf{48}, 1926 (1993);\\
\verb=http://gwdac.phys.gwu.edu/analysis/pd_analysis.html/=
%
\bibitem{Boschitz} E.T.~Boschitz \emph{et al.}, Phys.\ Rev.\ C
\textbf{6}, 457 (1972).
%
\bibitem{Katayama} N.~Katayama, F.~Sai, T.~Tsuboyama,
S.S.~Yamamoto, Nucl.\ Phys.\ A \textbf{438}, 685 (1985).
%
\bibitem{Glauber} R.J.~Glauber, in \textit{Lectures in Theoretical
Physics}, ed.\ W.E.~Brittin (Interscience, N.Y.\ 1959) vol.~1,
p.~315.
%
\bibitem{Komarov} V.I.~Komarov \emph{et al.}, Phys.\ Lett.\ B
  \textbf{553}. 179 (2003).
%
\bibitem{EDDA} B.~Lorentz \emph{et al.}, in \emph{Proceedings of the 9th
  European Particle Accelerator Conference, Lucerne, Switzerland, 2004}
  (EPS-AG CERN, Geneva, 2005), p.~1246.
%
\bibitem{Ohlsen} G.G.~Ohlsen, Rep.\ Prog.\ Phys.\ \textbf{35},
717 (1972).
%
\bibitem{Gomez} R.~Engels \emph{et al.}, AIP Conf.\ Proc.\ \textbf{980}, 161
  (2007).
%
\bibitem{Barbaro} M.B.~Barbaro, C.~Wilkin, J.\ Phys.\ G
\textbf{15}, L69 (1989).
%
\bibitem{Stein} H.J.~Stein \emph{et al.}, Phys.\ Rev.\ ST Accel.\ Beams, \textbf{11},
052801 (2008).
\end{thebibliography}
\end{document}